\documentclass[12pt,preprint]{aastex}
\usepackage{amsmath}
\usepackage{amssymb}
\usepackage{multirow}
\usepackage{gensymb}
\usepackage{graphicx}
\usepackage{float}
\usepackage{placeins}
\DeclareGraphicsExtensions{.jpg}
\usepackage{epstopdf}

\shorttitle{LSR1610$-$0040}
\shortauthors{Koren et al.}

\begin{document}
\title{The Low-Mass Astrometric Binary LSR1610$-$0040}
\author{Seth~C.~Koren\altaffilmark{1,3}, Cullen~H.~Blake\altaffilmark{1}, Conard C. Dahn\altaffilmark{2}, and Hugh C. Harris\altaffilmark{2}}

\altaffiltext{1}{Department of Physics \& Astronomy, University of Pennsylvania, 209 South 33rd Street, Philadelphia, PA 19104, USA}
\altaffiltext{2}{US Naval Observatory, Flagstaff Station, 10391 West Naval Observatory Road, Flagstaff, AZ 86001-8521, USA}
\altaffiltext{3}{Department of Physics, University of California, Santa Barbara, CA 93106, USA}
\begin{abstract}
Even though it was discovered more than a decade ago, LSR1610$-$0040 remains an enigma. This object has a peculiar spectrum that exhibits some features typically found in L subdwarfs, and others common in the spectra of more massive M dwarf stars. It is also a binary system with a known astrometric orbital solution. Given the available data, it remains a challenge to reconcile the observed properties of the combined light of LSR1610$-$0040AB with current theoretical models of low-mass stars and brown dwarfs. We present the results of a joint fit to both astrometric and radial velocity measurements of this unresolved, low-mass binary. We find that the photocentric orbit has a period $P = 633.0 \pm 1.7$ days, somewhat longer than previous results, with eccentricity of $e = 0.42 \pm 0.03$, and we estimate that the semi-major axis of the orbit of the primary is $a_1 \approx 0.32$ AU, consistent with previous results. While a complete characterization of the system is limited by our small number of radial velocity measurements, we establish a likely primary mass range of 0.09 - 0.10 $M_\odot$ from photometric and color-magnitude data. For a primary mass in this range, the secondary is constrained to be 0.06 - 0.075 $M_\odot$, making a negligible contribution to the total I-band luminosity. This effectively rules out the possibility of the secondary being a compact object such as an old, low-mass white dwarf. Based on our analysis, we predict a likely angular separation at apoapsis comparable to the resolution limits of current high-resolution imaging systems. Measuring the angular separation of the A \& B components would finally enable a full, unambiguous solution for the masses of the components of this system.
\end{abstract}


\maketitle

\section{\large Introduction} 
\vspace{2ex}
The discovery of the low mass, high proper motion object LSR1610$-$0040 (hereafter LSR1610) was first reported by \citet{lepine} based on Palomar Sky Survey photometry and follow-up red-optical spectroscopy. Due to its high galactocentric velocity ($> 200~\textrm{km\,s}^{-1}$), overall Spectral Energy Distribution (SED), and spectral features indicative of low metallicity, LSR1610 was assumed to be a member of an old stellar population. Given its optical and near-infrared colors and spectral features, it was clear that this object was cool and low in mass, but LSR1610 resisted clear classification on the M/L/T spectral sequence due to a number of peculiar features in its spectrum. Based primarily on spectral features common to L dwarfs, indications of a metal-poor atmosphere, and the significantly redder color than would be expected for an M subdwarf, \citet{lepine} proposed classifying LSR1610 as one of the first early-type L subdwarfs ever discovered.

Shortly thereafter, \citet{reiners} and \citet{cushing} presented new spectra in the red-optical and near-infrared, respectively. These papers came to the mutual conclusion that LSR1610 should be classified as an M-type star - though a ``schizophrenic'' one. \citet{cushing}, in particular, noted that the broad-band spectral energy distribution of LSR1610 was most consistent with an M dwarf spectrum, and that a number of the peculiar features (particularly those that suggested a subdwarf classification) could be explained if LSR1610 were just ``mildly metal-poor''. Still, there were some spectral features that this picture did not account for, such as Rb I lines that are typically present only in L-type stars, Al I lines that were stronger than expected for metal-poor M/L stars, and an ``unidentified, triangular-shaped'' absorption band around 0.9375~$\micron$, possibly due to TiO. By comparing their high-resolution optical to PHOENIX synthetic spectra, \citet{reiners} conclude that the optical spectrum of LSR1610 is reasonably well fit by a model with $T_{eff}=2800$~K and $[\rm{Fe}/\rm{H}]=-1$. Given the available spectral evidence, and the results of the analyses by \citet{cushing} and \citet{reiners}, \citet{dahn} proposed classifying LSR1610 as sd?M6pec.    

\citet{dahn} published a large set of astrometric measurements that provided a parallax for LSR1610 ($\pi=31.02\pm0.26$~mas), and also demonstrated the existence of an unresolved companion to LSR1610A. With an astrometric orbital solution based on the motion of the photocenter, which is the position of the combined light of LSR1610AB, \citet{dahn} measured the period, inclination, and eccentricity of this system. Based on their direct distance measurement and galactocentric velocity, \cite{dahn} also concluded that, kinematically, LSR1610AB was likely a member of the galactic halo population. However, given the typical age of halo objects (10~Gyr), the metallicity of LSR1610 might be expected to be lower than $[\rm{Fe}/\rm{H}]=-1$. \citet{dahn} also pointed out that while the near infrared (NIR) colors of LSR1610 are roughly consistent with available isochrones, LSR1610 is actually a significant outlier in the V-I vs. B-V color-color space, more than 1.5 magnitudes redder in B-V than other known M dwarfs and subdwarfs. Finally, \citet{dahn} proposed the intriguing hypothesis that the observed properties of LSR1610 could be the result of mass transfer. If a giant companion deposited material onto the dwarf star LSR1610A, this could explain the enhanced Al abundance and the decreased flux at blue wavelengths, where flux would be suppressed by molecular absorption bands resulting from the accreted material. In this scenario, the companion LSR1610B would now be an old white dwarf, or possibly a neutron star, the remnant of the Red Giant (RG) or Asymptotic Giant Branch (AGB) star that lost material to LSR1610A. However, given models of old, low-mass Helium core white dwarfs \citep{althaus}, at a distance of just 32.3 pc this companion should be directly detectable in the UV and optical in SDSS photometry. 

Directly measuring the masses of the components of LSR1610AB, or determining their luminosity ratio, requires additional observations. \citet{dahn} compared the fitted photocentric orbit to a radial velocity measurement, but did not have enough radial velocity measurements to incorporate these directly into the orbital fitting, though the available radial velocity data were consistent with LSR1610B being a very low mass, low-luminosity object such as a brown dwarf. \citet{blake} obtained additional radial velocity measurements of this system, and performed a fit for the radial velocity orbital solution assuming the period of the photocentric orbit from \citet{dahn}. Measuring the radial velocity semi-amplitude allowed for the testing of a range of hypotheses about the possible component masses and luminosities, as well as additional comparisons to theoretical models of low-mass stars and brown dwarfs. While the large variations in the radial velocity of LSR1610AB observed by \citet{blake} were fully consistent with the photocentric orbital period, the derived radial velocity semi-amplitude appeared inconsistent with plausible scenarios for the masses and luminosities of the A and B components. In an effort to resolve these inconsistencies, here we present a joint analysis of astrometric and radial velocity observations of LSR1610 and place strong constraints on the possible masses and luminosities of both LSR1610 A and B. 

\section{\large Data}

Our analysis combines NIR radial velocity measurements published in \citet{blake} with optical differential astrometric measurements from \citet{dahn}, supplemented with additional astrometric measurements from the USNO parallax program. In Table \ref{data} we give a summary of the basic characteristics of these two data sets. The radial velocity measurements were derived from spectra obtained with the NIRSPEC instrument on the Keck II telescope. NIRSPEC is a high-resolution, cross-dispersed NIR echelle spectrograph \citep{mclean1998}. A detailed description of the observational procedures used to obtain our radial velocity measurements can be found in \citet{blake}. In total, four epochs of NIRSEPC observations were gathered spanning 4.2 years. Observations were made around $2.3~\micron$, where there is a strong CO bandhead feature in the spectra of cool stars, affording typical radial velocity precision of $200$~m~s$^{-1}$ using telluric methane lines in Earth's atmosphere as a wavelength reference. Our astrometric observations come from the TEK2K CCD Camera on the 1.55~m Strand Astrometric Reflector at the Flagstaff Station of the US Naval Observatory and were gathered as part of the USNO faint-star parallax program. The observational procedures followed were summarized in \citet{dahn2}. The astrometric measurements were made in the I band. We assume uniform estimates for the errors of the RA and DEC astrometric positions at each epoch, with slightly larger error estimates attached to the DEC measurements: 2.25 mas in RA and 3.25 mas in DEC. These errors are based on the average residuals in the astrometric fits to data from a large sample of stars. In addition to those measurements already published in \citet{dahn}, we included more recent astrometric data from the ongoing USNO parallax program to extend the total time baseline to approximately 10 years.

\section{\large Methods}

In order to improve constraints on the physical properties of LSR1610AB, we carried out a joint fit to the combined set of astrometric and radial velocity measurements. Starting from the \citet{dahn} orbital solution, we used Markov chain Monte Carlo (MCMC) methods to obtain an updated solution for the orbital and system parameters for LSR1610AB incorporating the expanded set of astrometric measurements. MCMC methods sample the posterior probability density function (PDF) of a set of model parameters, ${\bf\Xi}$, effectively evaluating the likelihood of those parameters given the data. This requires the specification of a likelihood function, $L[{\bf\Xi}]$.  The MCMC algorithm then compares $L[{\bf\Xi}]$ for a proposed set of trial parameters to $L[{\bf\Xi}]$ for the previously accepted set of parameters and determines an acceptance probability for the proposed parameters using only the ratio of likelihoods \citep{ford,foreman}. The likelihood $L[{\bf\Xi}]$ of our model is:

\begin{equation}
L[{\bf\Xi}] = \kappa e^{-\frac{1}{2}\chi^{2}[{\bf\Xi}]} \label{likelihood}
\end{equation}
\begin{align}
\chi^{2}[{\bf\Xi}] &= \sum\limits_{j}^{N_{RV}} \left(\frac{RV_{obs,j} - RV_{model,j}[t,{\bf\Xi}]}{\sigma_j}\right)^{2} \nonumber \\
&+ \sum\limits_{k}^{N_{astro}} \left(\frac{RA_{obs,k} - RA_{model,k}[t,{\bf\Xi}]}{\sigma_{RA}}\right)^{2} \nonumber \\
&+ \sum\limits_{k}^{N_{astro}} \left(\frac{DEC_{obs,k} - DEC_{model,k}[t,{\bf\Xi}]}{\sigma_{DEC}}\right)^{2} \label{chi}
\end{align}

\noindent where $\chi^{2}$ is the familiar weighted sum of squares statistic, $RV_{obs,i}$, $RA_{obs,i}$, and $DEC_{obs,i}$ represent single radial velocity or astrometric measurements, $\sigma_i$ are the associated uncertainties on those measurements, and $x_{model,i}$ is the Keplerian orbit model given the parameter vector ${\bf\Xi}$. The constant $\kappa$ is a normalization that can be ignored, since MCMC methods evaluate only $ {L[{\bf\Xi}_1]}/{L[{\bf\Xi}_2]} $ \citep{ford}. The likelihood function $L[{\bf\Xi}]$ multiplied by the priors on the parameters is proportional to the PDF of the parameters given the observations, $P[{\bf\Xi}]$. We used priors only to enforce definitions of parameters and mathematical equivalence ($\aleph \geq 0; \ P > 0; \ 0 \leq e < 1; \ \varpi > 0; \ 0 \leq i < \pi$ \ - see below for parameter definitions). We used the $emcee$ package to carry out our MCMC analysis. $emcee$ provides a sampler that is affine invariant and utilizes a large ensemble of ``walkers'' to efficiently sample from complex posterior distributions \citep{foreman}. 

Calculating $L[{\bf\Xi}]$ requires evaluating the model $x_{model,k}$ a large number of times given trial sets of free parameters, ${\bf\Xi}$, and the specific times of the actual observations. We follow the definitions of the Keplerian astrometric and radial velocity orbits as described in \citet{wright}. With radial velocity data, the reflex motion of the primary star along our line of sight is measured. If we denote the mass of the primary star (LSR1610A) in the binary system as $m_1$, with $a_1$ the semi-major axis of its Keplerian orbit around the system barycenter, then modeling the radial velocity data requires six free parameters:
\begin{itemize}
\item $K_1$ - The semi-amplitude of the oscillation of the primary star's radial velocity
\item $\omega$ - The argument of periastron of the primary star's orbit
\item $e$ - The eccentricity of the Keplerian orbit
\item $t_p$ - The time of periastron passage
\item $P$ - The orbital period of the system
\item $\gamma$ - A constant radial velocity offset
\end{itemize}
The first five of these parameters determine the Keplerian orbit of the system; the last is needed to fit the data and gives us information on the systematic space motion of the system, but is unimportant for characterizing the orbital motion of the binary system.

With astrometric observations of an unresolved binary system, the reflex motion of the system photocenter in the transverse direction is measured on the sky. Modeling the astrometric data requires twelve free parameters:
\begin{itemize}
\item $\omega$ - The argument of periastron of the primary star's orbit
\item $e$ - The eccentricity of the Keplerian orbit
\item $t_p$ - The time of periastron passage
\item $P$ - The orbital period of the system
\item $\Omega$ - The longitude of the ascending node of the secondary star's orbit
\item $i$ - The inclination angle of the system
\item $\aleph$ - The semi-major axis of the orbit of the photocenter in angular units
\item $\mu_\alpha$ - The proper motion of the system in Right Ascension 
\item $\mu_\delta$ - The proper motion of the system in Declination 
\item $\varpi$ - The parallax of the system
\item $\alpha_0$ - The fiducial position of the system in Right Ascension 
\item $\delta_0$ - The fiducial position of the system in Declination 
\end{itemize}
In our case, the proper motion and fiducial position parameters are unimportant for characterizing the orbital motion of the system, and the MCMC method allows us to easily marginalize them.

When both radial velocity and astrometric data are available, a joint fit may be performed. As the two models share four parameters ($P$,$\omega$,$e$,$t_p$), the joint fit has fourteen free parameters. From a given parameter vector ${\bf\Xi}$ = [$P, \omega, e, t_p, K_1, \Omega, i, \aleph, \varpi, \gamma, \mu_\alpha, \mu_\delta, \alpha_0, \delta_0$], model values can can be generated for use in Equation \ref{chi} following \citet{wright}:

\begin{equation}
RV{_{model}}[t,{\bf\Xi}] = K_1\left[\cos(\omega + \nu(t)) + e\cos\omega\right] + \gamma \label{rv}
\end{equation}
\begin{align}
RA{_{model}}[t,{\bf\Xi}] = &BX(t) + GY(t) +  \mu_{\alpha}(t - t_0)\nonumber \\
&+ \varpi\Pi_{\alpha}(t) + \alpha_0 \cos\delta \label{ra}
\end{align}
\begin{align}
DEC{_{model}}[t,{\bf\Xi}] = &AX(t) + FY(t) + \mu_{\delta}(t - t_0) \nonumber \\
&+ \varpi\Pi_{\delta}(t) + \delta_0 \label{dec}
\end{align}

\noindent where $\nu(t)$ is the true anomaly; $A$, $B$, $F$, and $G$ are the Thiele-Innes constants; $X(t)$ and $Y(t)$ are the elliptical rectangular coordinates, and $\Pi_{\alpha}(t)$ and $\Pi_{\delta}(t)$ are the astrometric displacements due to parallax in the RA and DEC directions, respectively. 

The true anomaly is the angle between the argument of periastron and the current position of the star as measured from the main focus of the elliptical orbit, and can be calculated numerically using Newton's method (or a similar method) from the following definitions:
\begin{equation}
\tan\frac{\nu(t)}{2} = \sqrt{\frac{1 + e}{1 - e}}\tan\frac{E(t)}{2} \label{true}
\end{equation}
\begin{equation}
E(t) - e\sin E(t) = M(t) \label{eccentric}
\end{equation}
\begin{equation}
M(t) = \frac{2\pi(t - t_p)}{P} \ \bmod{\ 2\pi} \label{mean}
\end{equation}
\noindent where the mean anomaly $M(t)$ is calculated from $[t,{\bf\Xi}]$, and the step $M(t) \rightarrow E(t)$ requires a numerical solution for the eccentric anomaly.
The Thiele-Innes constants are projected rectangular equatorial coordinates \citep{van}, and are calculated from ${\bf\Xi}$:
\begin{alignat}{3}
A \ = \ \aleph(&+\cos\Omega\cos\omega &- \sin\Omega\sin\omega\cos i) \label{TIA} \\ 
B \ = \ \aleph(&+\sin\Omega\cos\omega &+ \cos\Omega\sin\omega\cos i) \label{TIB} \\ 
F \ = \ \aleph(&-\cos\Omega\sin\omega &- \sin\Omega\cos\omega\cos i) \label{TIF} \\ 
G \ = \ \aleph(&-\sin\Omega\sin\omega &+ \cos\Omega\cos\omega\cos i) \label{TIG}
\end{alignat}
The elliptical rectangular coordinates X(t) and Y(t) are defined as:
\begin{align}
X(t) \ =& \ \cos E(t) - e \label{X} \\
Y(t) \ =& \ \sqrt{1 - e^{2}}\sin E(t) \label{Y}
\end{align}

\noindent where $E(t)$ is the eccentric anomaly defined in Equation \ref{eccentric}.
Finally, the astrometric displacements due to parallax are:
\begin{align}
\Pi_{\alpha}(t) =&  \ r_x(t)\sin\alpha - r_y(t)\cos\alpha \label{pi_ra}\\ 
\Pi_{\delta}(t) =& \  (r_x(t)\cos\alpha + r_y(t)\sin\alpha)\sin\delta \nonumber \\
&- r_z(t)\cos\delta \label{pi_dec}
\end{align}
Here, $\alpha$ and $\delta$ are the fiducial RA and Dec of the system, which are taken to be $[\alpha = 16^{\textrm{h}}10^{\textrm{m}}29.0^{\textrm{s}},\delta = -00\degree 40'53.0"]$. The Cartesian components in equatorial coordinates of the position vector $\vec r$ from the Solar System barycenter to the center of the Earth are  $r_x(t)$, $r_y(t)$, and $r_z(t)$ (expressed in AU if $\varpi$ is in arcseconds). These values can be obtained by querying the NASA Jet Propulsion Laboratory's online HORIZONS system for periodic ephemeris values spanning the time period of the astrometric data, and linearly interpolating to find exact values corresponding to the epochs of the astrometric observations\footnote{http://ssd.jpl.nasa.gov/?horizons}.

If the LSR1610 system were composed of a primary, light-emitting star and a secondary, dark companion (dark in the I-band where the astrometric measurements were made), the photocentric semi-major axis, $\aleph$, in the Thiele-Innes constants definitions (Equations \ref{TIA} - \ref{TIG}) would be equivalent to $a_1\varpi$, the semi-major axis of the orbit of the primary star in angular units. However, it is possible that the components of LSR1610AB have comparable luminosities in the I band, in which case the measured $\aleph$ will not be equal to  $a_1\varpi$. We can define $\beta_{I}$, the luminosity ratio of the primary and secondary stars \citep{dahn}:
\begin{equation}
\beta_{I} = \frac{\iota_2}{\iota_1 + \iota_2} \label{beta} 
\end{equation}
\\ \noindent where $\iota_1$ is the luminosity of the primary in the I band and $\iota_2$ is the luminosity of the secondary in the I band.
We can then give the relationship between $\aleph$ and $a_1\varpi$:
\begin{equation}
a_1 = \frac{\aleph}{\varpi}\textrm{AU}\left[1 - \beta_{I}\left(\frac{m_1 + m_2}{m_2}\right)\right]^{-1}  \label{aaleph}
\end{equation}
\noindent where $m_1$ is the mass of the primary and $m_2$ is the mass of the secondary. Unfortunately, with an \textit{unresolved} binary the degeneracy between $\beta_{I}$, $m_1$, and $m_2$ cannot be broken, even with both astrometric and radial velocity data.

Given that the components have elliptical orbits, and from the conservation of momentum, we have:
\begin{equation}
a_1 = K_1 \frac{P\sqrt{1 - e^{2}}}{2\pi\sin i} \label{kepler}
\end{equation}
If we assume that the radial velocity semi-amplitude we measure corresponds to the reflex motion of the brighter (in K band), more massive, primary component LSR1610A, then from Kepler's third law we get the mass function for a spectroscopic binary system:
\begin{equation}
\frac{m_{2}^{3}\sin i^{3}}{(m_1 + m_2)^{2}} =  K_1^{3}\frac{P(1 - e^{2})^{3/2}}{2\pi G} \label{mass}
\end{equation}
The inclination, $i$, is measured from the astrometric orbit, and the eccentricity, $e$, and orbital period, $P$, are measured from both the astrometric and radial velocity orbits, with $G$ being the gravitational constant. This gives us three equations (Equations \ref{aaleph}, \ref{kepler}, \ref{mass}) in four unknowns ($a_{1}$,$\beta_{I}$,$m_1$,$m_2$). We can reorganize these equations to better express the unknown quantities, and define new variables that are functions of the parameters we actually measure:

\begin{equation}
\beta_{I}\left(\frac{m_1 + m_2}{m_2}\right) = 1 - \frac{\aleph}{\varpi}\textrm{AU}\frac{1}{a_1} \equiv X_1 \label{x1}
\end{equation}
\begin{equation}
\frac{m_2^{3}}{(m_1 + m_2)^{2}} = K_1^{3} \frac{P(1 - e^{2})^{3/2}}{2\pi G}\frac{1}{\sin i^{3}} \equiv X_2 \label{x2}
\end{equation}
By using external constraints on $\beta_{I}$, $m_1$, and $m_2$, derived from  broad-band photometry and our knowledge of stellar evolution, we can place strong constraints on the components of LSR1610 given the available data. 

\section{\large Results}

We applied the \textit{emcee} algorithm iteratively to the combined astrometric and radial velocity data set to ensure convergence, and used a large distribution of walkers initiated spanning a wide parameter space. We used $emcee$'s $acor$ package to determine the necessary number of burn-in steps to be used with the MCMC algorithm. The autocorrelation is a measure of the number of proposed steps that are necessary to produce independent samples of the PDFs.  We threw out ten times this number of steps from each chain to ensure that the PDFs were being sampled effectively. The mean acceptance fraction of the chains (number of steps accepted to number of steps proposed) was 0.36, which is well within its optimal range (0.2 - 0.5; \citealt{foreman}).

For each MCMC iteration, we initialized 10 chains with uniformly random parameter vectors $p0$ all within a proposal distribution width $\Delta$ of some starting parameter vector ${\bf\Xi}_0$. We distributed the walkers for each chain around each chain's $p0$ in a ''sample ball" with the same width $\Delta$.  For the first iteration, we used the best fit parameters from  \citet{dahn}, supplemented with guesses for those parameters the astrometric data alone do not constrain. For each successive MCMC iteration, we set ${\bf\Xi}_0$ to the vector of means of the PDFs for each parameter from the chains in the previous iteration. This iterative process was halted when the PDFs returned by successive iterations had means which differed by less than a tenth of a standard deviation, indicating convergence of the Markov chain to the posterior probability distributions. 

\begin{figure*}[ht]
\centering
\includegraphics[width=\textwidth]{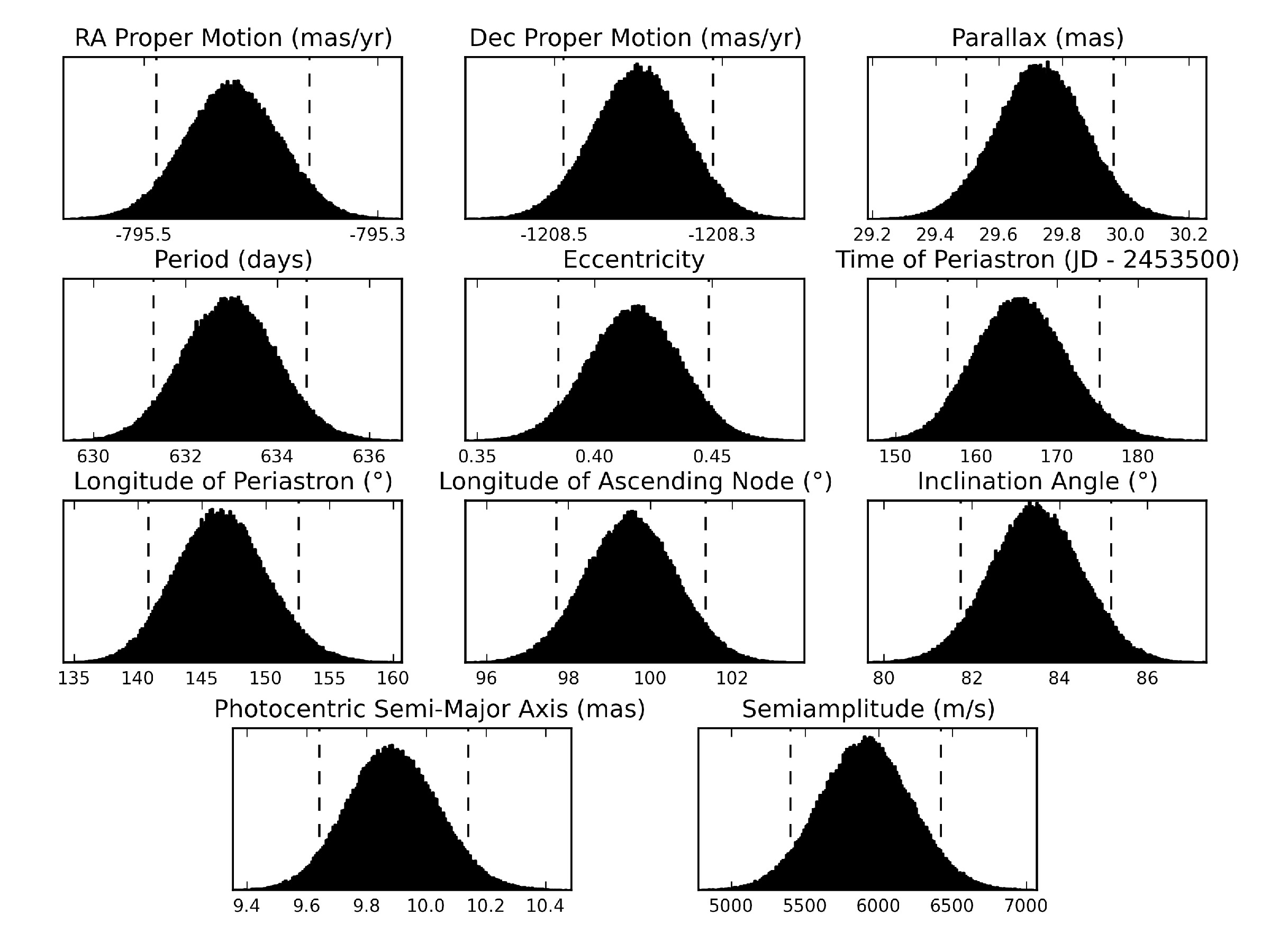}
\caption{Posterior probability distributions from the final Markov chain. Dashed lines demarcate 90\% confidence intervals. The distribution for parallax appears as returned by the chain, before the correction has been applied as described in Section 4. \label{distributions}}
\end{figure*}

\begin{figure*}[ht]
\centering
\includegraphics[width=\textwidth]{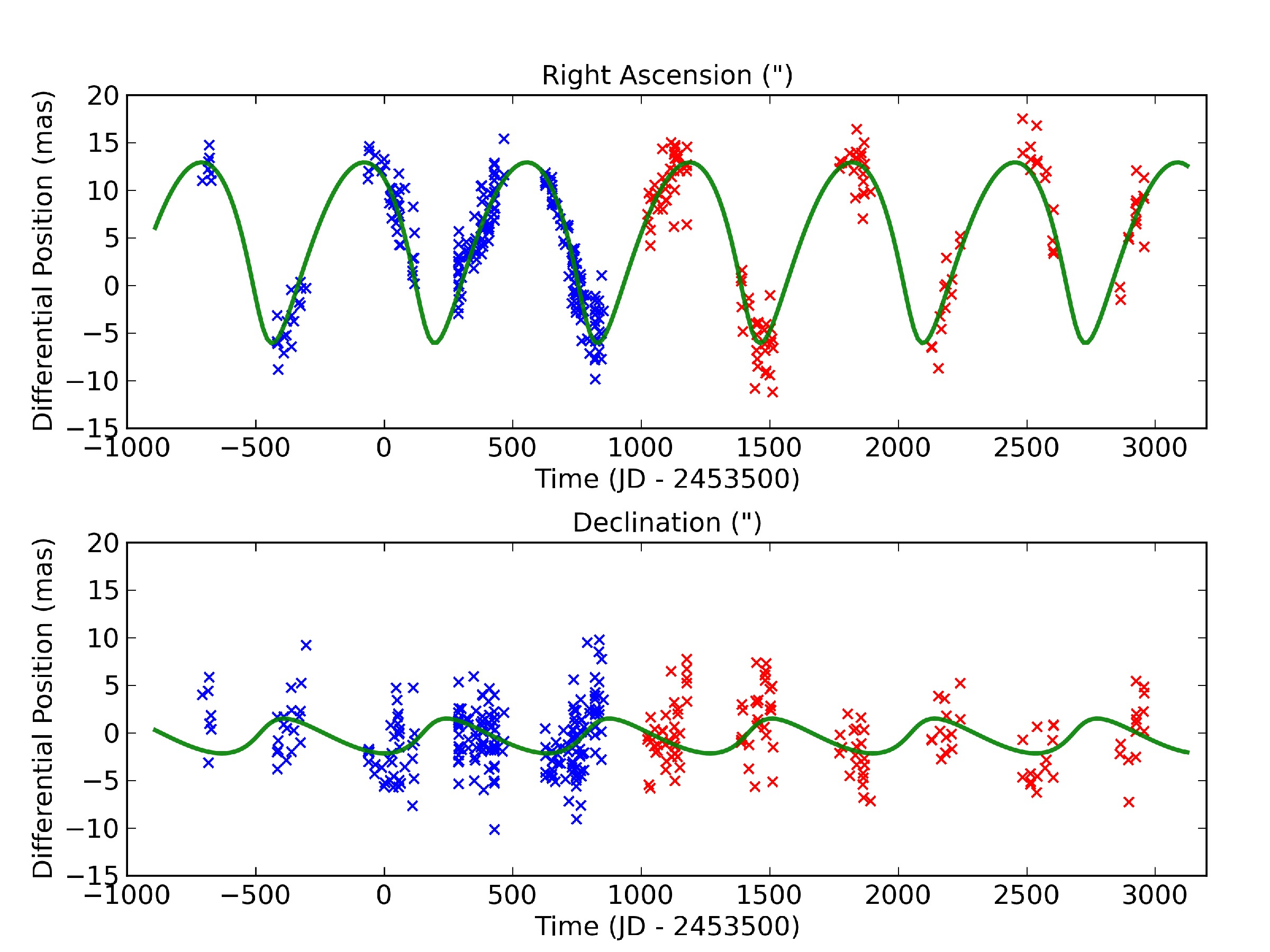}
\caption{Right Ascension and Declination data and model from the best-fit parameters (solid, green line). Parallactic and proper motion have been removed. Data previously published in Dahn et al. (2008) are in blue, new data are in red. \label{radecmodel}}
\end{figure*}

\begin{figure*}[ht]
\centering
\includegraphics[width=\textwidth]{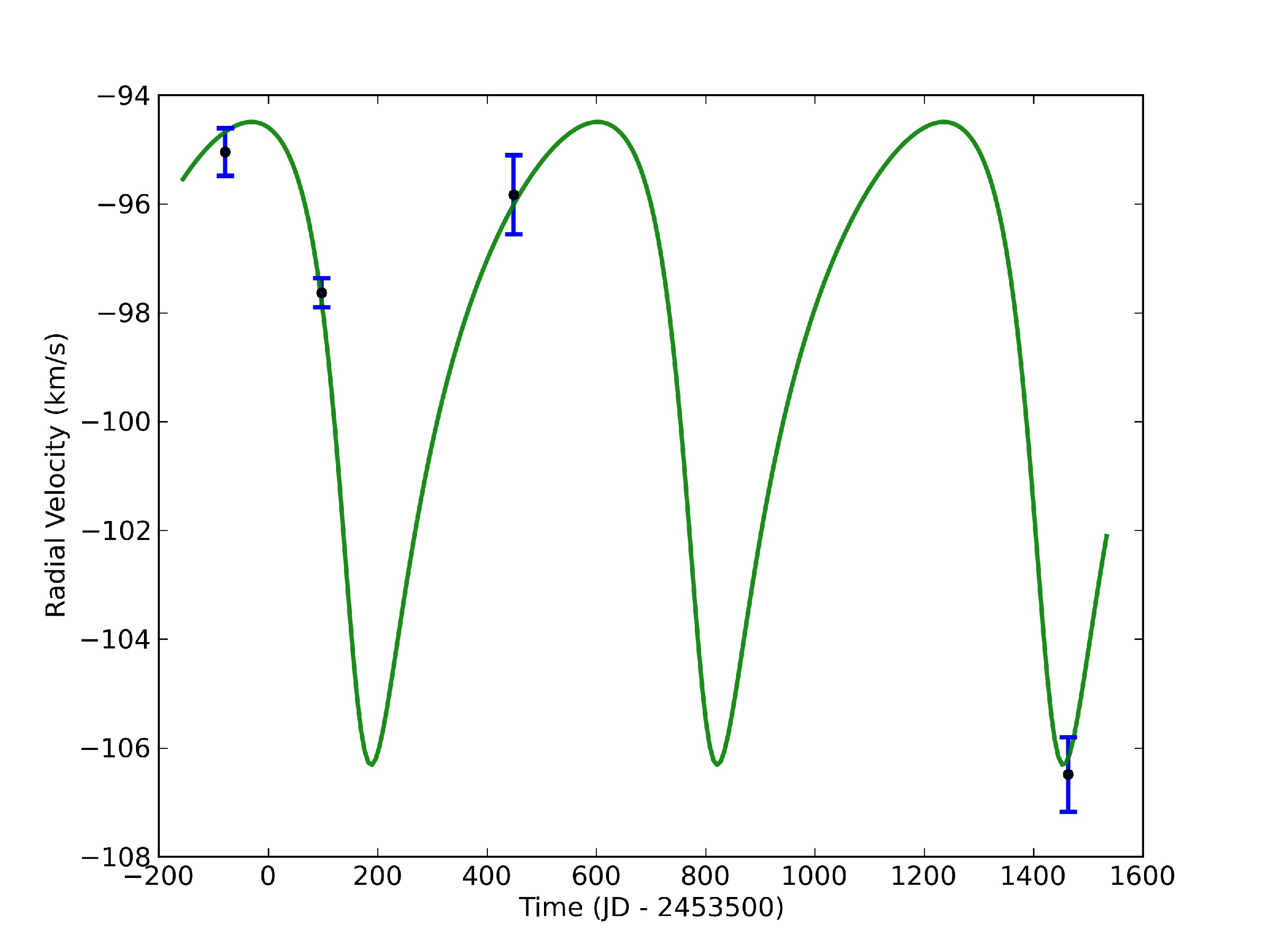}
\caption{Radial velocity data and model from the best-fit parameters (solid, green line). Data from Blake et al. (2010). \label{rvmodel}}
\end{figure*}

\begin{figure*}[ht]
\centering
\includegraphics[width=\textwidth]{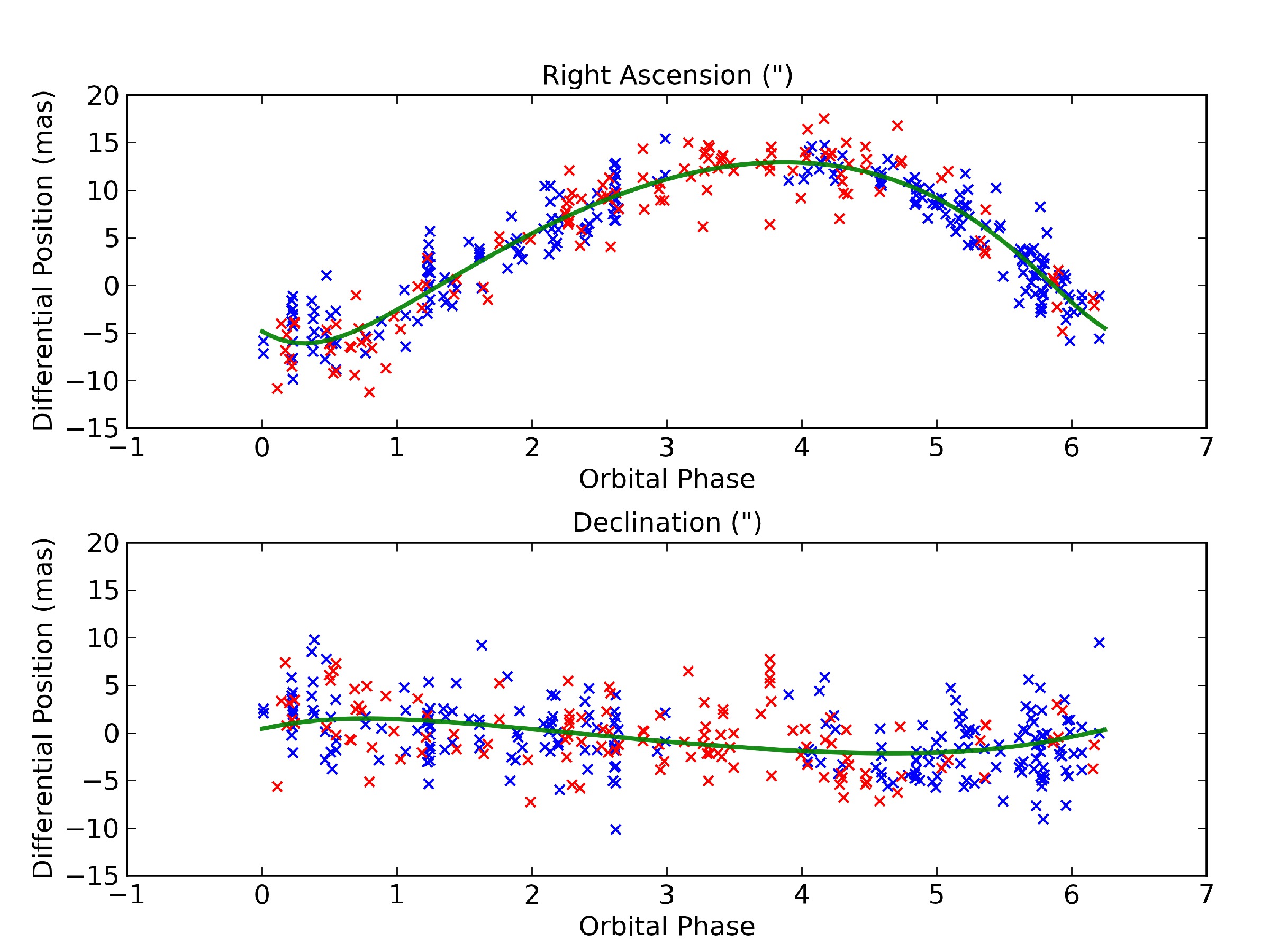}
\caption{Right Ascension and Declination data and best-fit model (solid, green line) by orbital phase in radians. Parallactic and proper motion have been removed. Data previously published in Dahn et al. (2008) are in blue, new data are in red. \label{radecmodelmod}}
\end{figure*}

\begin{figure*}[ht]
\centering
\includegraphics[width=\textwidth]{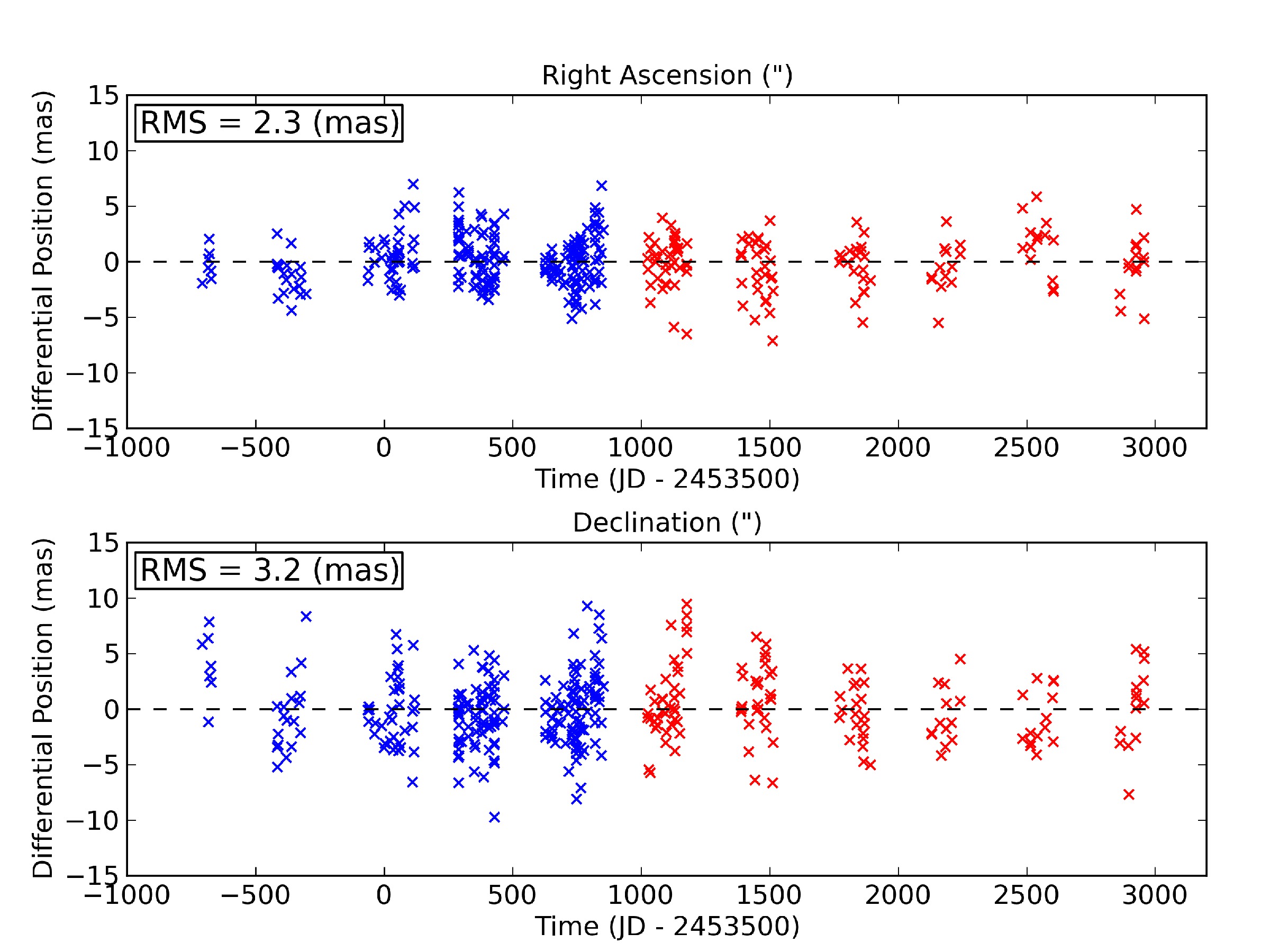}
\caption{Astrometric residuals from best-fit model. Data previously published in Dahn et al. (2008) are in blue, new data are in red. \label{radecres}}
\end{figure*}

\begin{figure*}[ht]
\centering
\includegraphics[width=\textwidth]{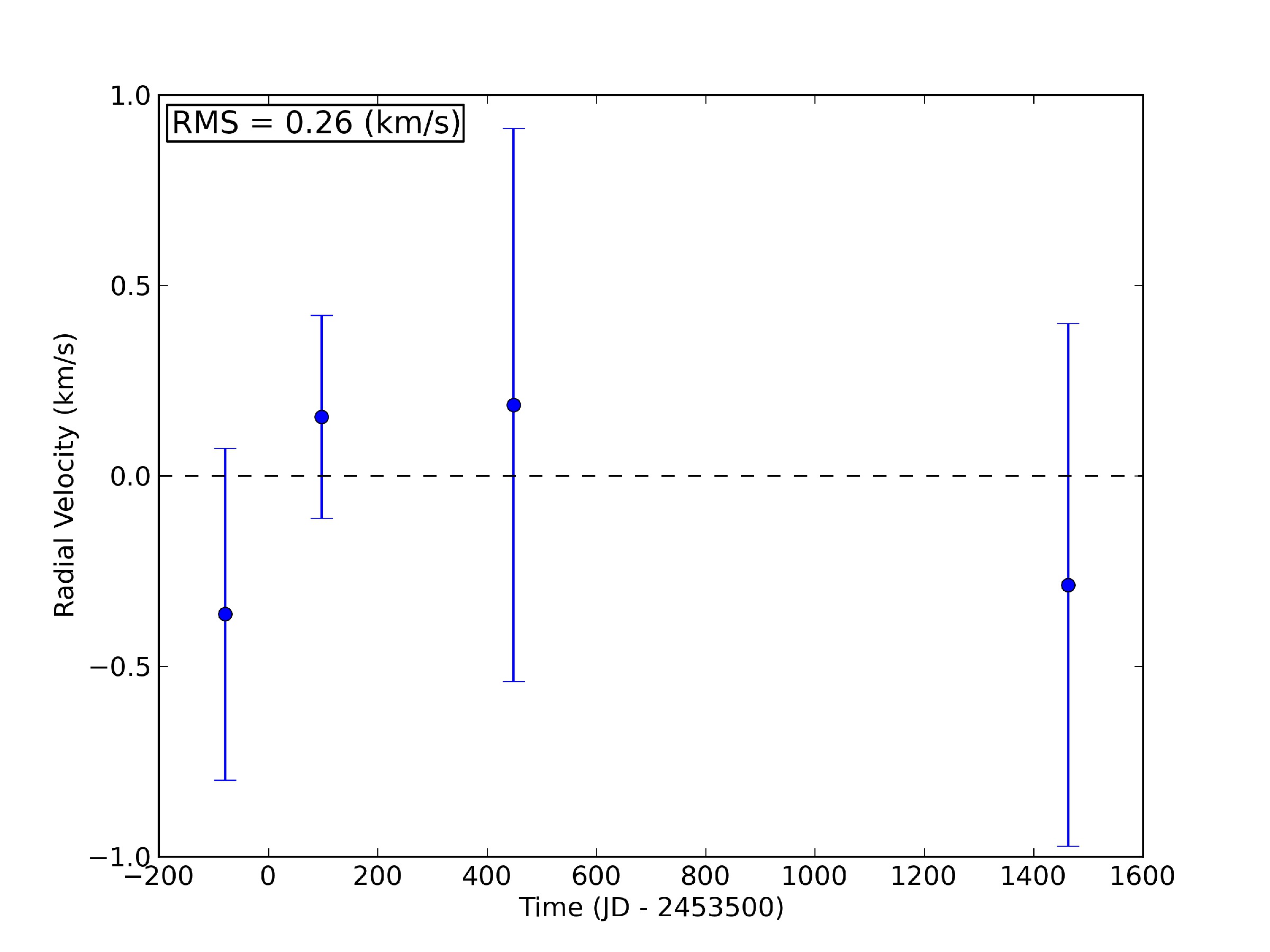}
\caption{Radial velocity residuals from best-fit model. \label{rvres}}
\end{figure*}

The final PDFs from the 9th iteration appear in Figure \ref{distributions}. From these, we assumed the simple mean of each chain, after discarding the burn in steps, as the maximum likelihood value for the parameter. Both astrometric and radial velocity models constructed from these best fit values are shown in Figures \ref{radecmodel} - \ref{radecmodelmod} and the residuals of these models appear in Figures \ref{radecres} and \ref{rvres}. We calculated the 90\% confidence interval on each parameter by sorting the chains and taking the 5th percentile and 95th percentile values. The point estimates and the 90\% confidence intervals on each parameter appear in Table \ref{parameters}. Goodness of fit information is given in Table \ref{goodness}.

We added a correction of $\Delta\pi=$+1.00 $\pm$ 0.15 mas to each sample of the parallax PDF to translate from relative parallax to absolute parallax \citep{dahn}. This correction brought our point estimate and confidence interval on parallax from 29.73 $\pm$ 0.23 mas to an absolute parallax of 30.73 $\pm$ 0.34 mas, fully consistent with the results from \citet{dahn} but somewhat lower than the absolute parallax reported by \citet{schilbach}. We note that the analysis presented in \citet{schilbach} does not include the astrometric perturbation caused by LSR1610B. As a consistency check, we ran an astrometric solution \textit{without} orbital motion using our data covering the same time period as the \citet{schilbach} observations and found an absolute parallax of $32.53\pm0.5$~mas, fully consistent with the \citet{schilbach} result. Any systematic biases in the relative to absolute parallax correction will result in systematic bias in both the distance and derived absolute magnitudes for LSR1610AB. Experience with USNO observations of many stars over many years indicates that a correction as large as 2.0 mas is strongly ruled out for LSR1610AB. This means that the impact of a biased $\Delta\pi$ on our subsequent analyses will be small. Internal consistency tests demonstrating the absolute accuracy of the USNO astrometric data carried out by observing a sample of quasars will be described in a future work.

\begin{figure*}[ht]
\centering
\includegraphics[width=.9\textwidth]{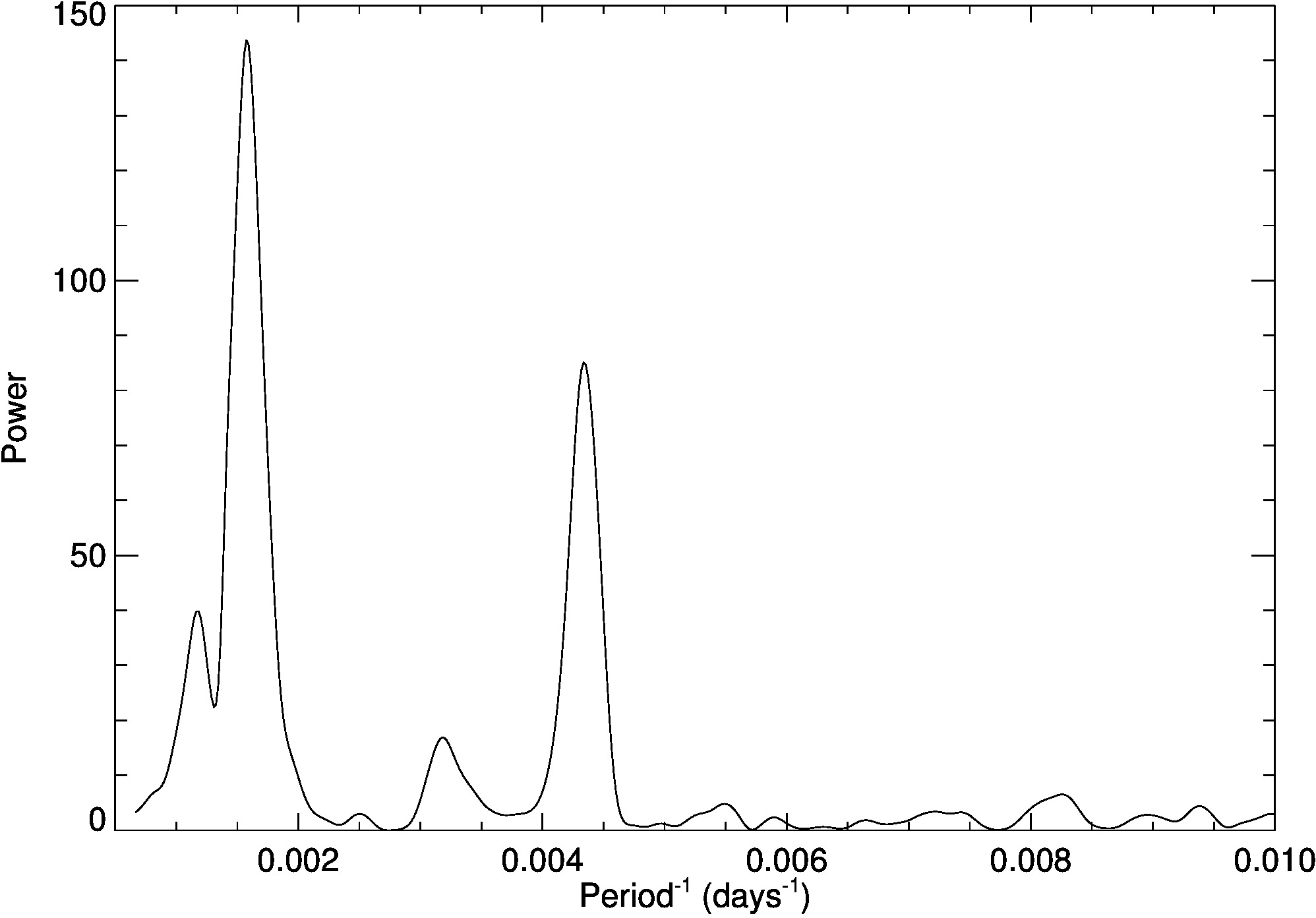}
\caption{Lomb-Scargle periodogram of the astrometric residuals after subtracting proper motion and parallax. The two peaks are at 634.9 and 230.6 days. \label{period}}
\end{figure*}

We used a Lomb-Scargle periodogram of the astrometric data, after removal of the proper motion and parallax, to check for possible alternative orbital periods that might be well outside our MCMC proposal distribution width $\Delta$ (see Figure \ref{period}). While the strongest periodogram peak coincided with the period reported by \citet{dahn}, a smaller peak at 230.6 days was also apparent. This peak is most likely an alias of the primary peak and the one-year sampling of the astrometric data. However, we attempted a series of fits using an initial ${\bf\Xi}_0$ with a period centered on this smaller secondary peak, but the resulting $\chi^{2}$ values were much worse than those incorporating the longer period of 633 days.

\begin{figure*}[ht]
\centering
\includegraphics[width=\textwidth]{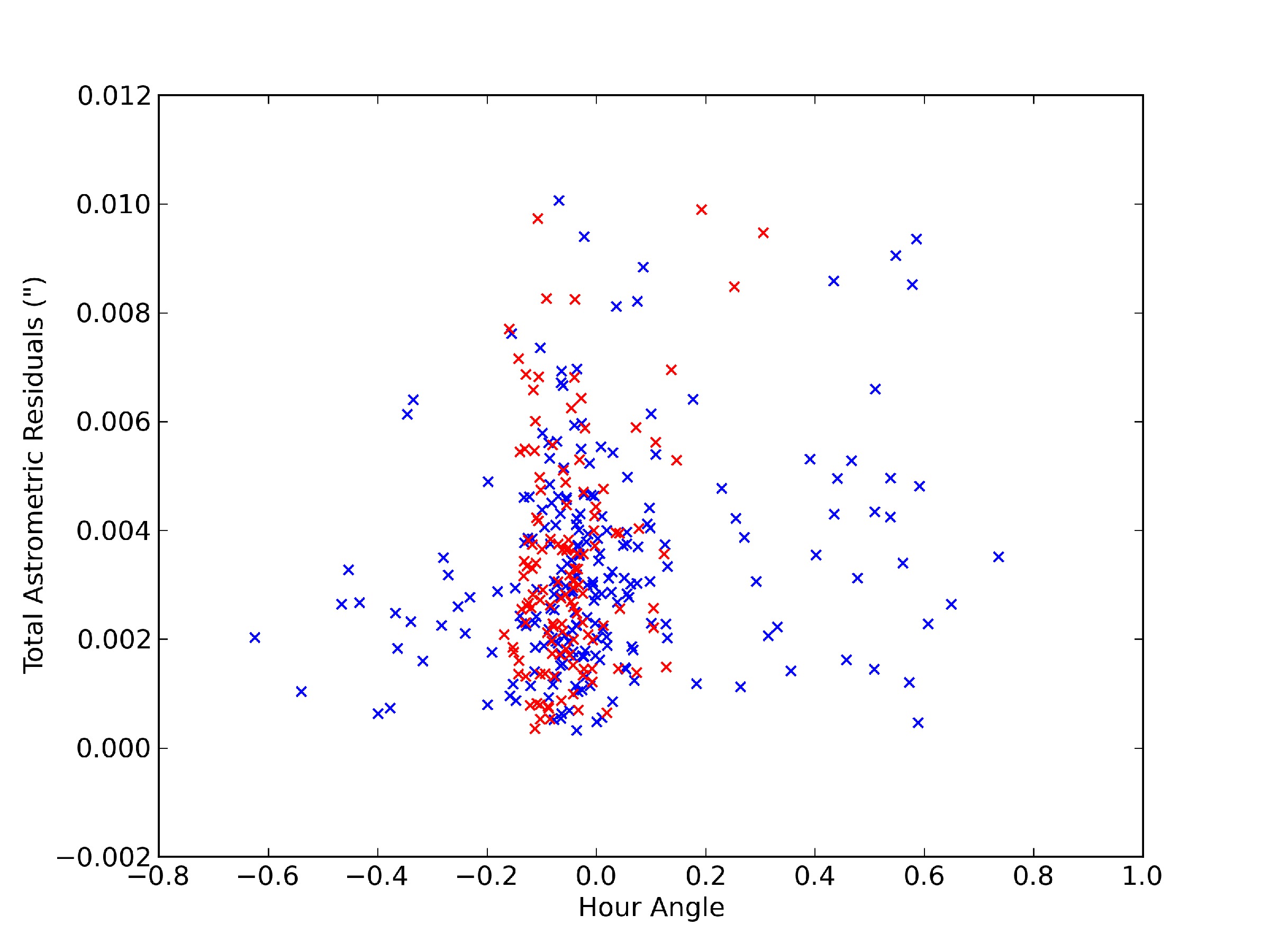}
\caption{Right Ascension and Declination residuals summed in quadrature against the hour angle at which each observation was made. Data previously published in \citet{dahn} are in blue, new data are in red. \label{hares}}
\end{figure*}

\begin{figure*}[ht]
\centering
\includegraphics[width=\textwidth]{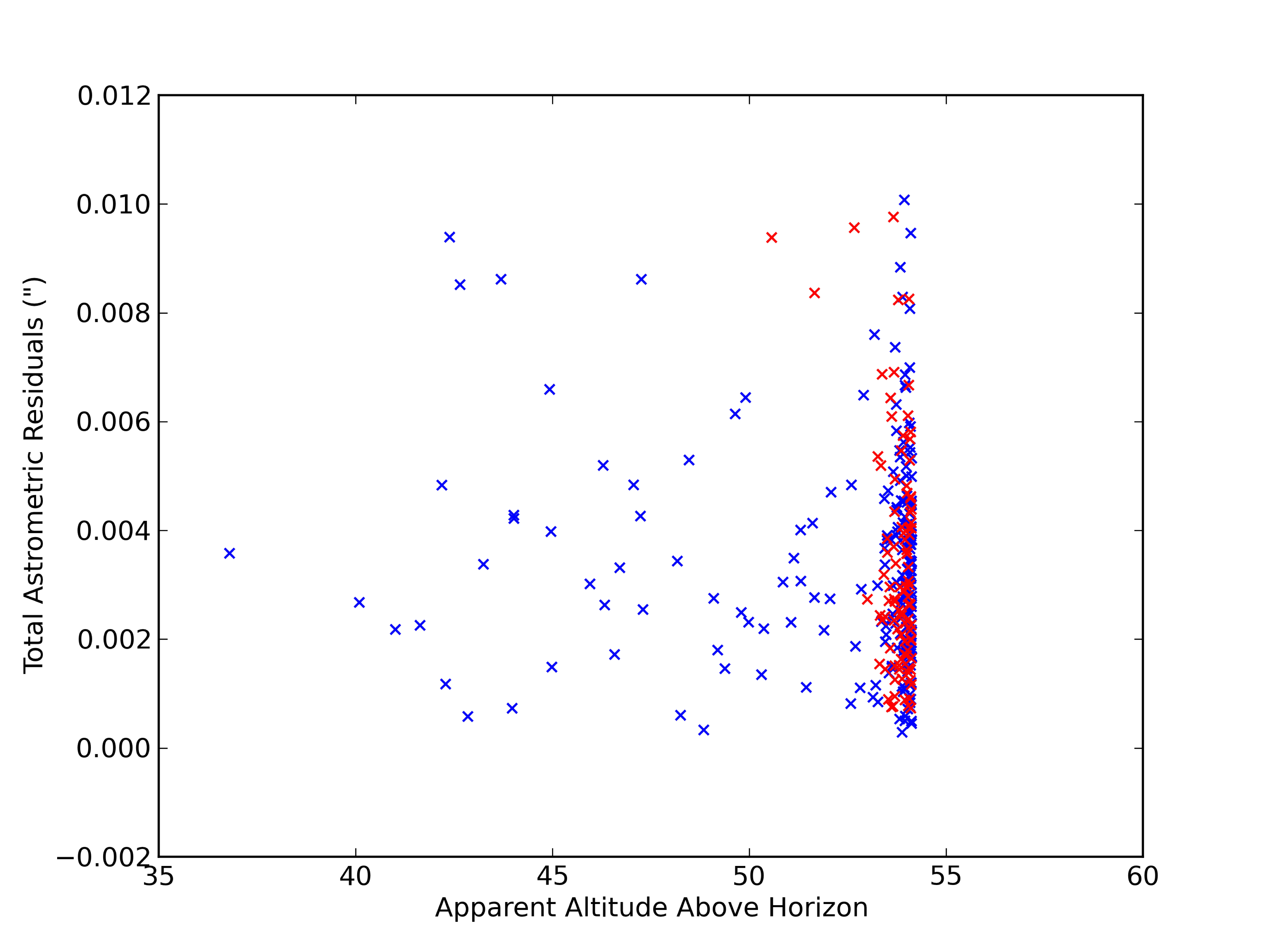}
\caption{Right Ascension and Declination residuals summed in quadrature against the altitude at which each observation was made. Data previously published in \citet{dahn} are in blue, new data are in red. \label{ares}}
\end{figure*}

We also attempted a fit that included additional parameters for a second, long period companion to LSR1610A. We found that the additional orbit did not improve the fit significantly, given the measurement uncertainties and the possibility of unknown covariance between measurements. Higher-precision astrometry would be necessary to rule out the possibility of a third component. We also checked the astrometric residuals for correlation with the hour angle of the observations and altitude at the time of the measurement, but we did not find statistically significant evidence for any correlations with these external parameters (see Figures \ref{hares} and \ref{ares}).

Using the results of the MCMC, we can derive a PDF for $X_2$ by sampling from the PDFs of the necessary parameters. This gives us $X_2 = 1.14^{+0.33}_{-0.27} \times 10^{-2} M_\odot$ to 90\% confidence. Using $X_2$, for given $m_1$ values we can now find a PDF for $m_2$ by numerically solving:

\begin{equation}
m_2^{3/2} - m_2\sqrt{X_2} - m_1\sqrt{X_2} = 0 \label{massx2}
\end{equation}

\noindent which follows from Equation \ref{x2}. We explore a range of possible $m_1$ values for which to calculate $m_2$ values given Equations \ref{x2} and \ref{massx2}. We start at $0.08 M_\odot$ because we assume the primary is hydrogen-burning, and limit ourselves to $m_{1}<0.3 M_\odot$ based on the known absolute magnitudes of the system. See Table \ref{masses} for a sample of calculated values of $m_2$ as a function of $m_1$.

By combining our two expressions for $a_1$ from Equations \ref{aaleph} and \ref{kepler} and the fact that $\beta_{I} \geq 0$ by definition, we can get a lower bound on the value of $K_1$ primarily based on quantities measured with the astrometric data:

\begin{equation}
K_1 \geq \frac{2\pi\sin i}{P\sqrt{1 - e^{2}}}\frac{\aleph}{\varpi}\textrm{AU} \equiv K_{min}
\end{equation}

\begin{figure*}[ht]
\centering
\includegraphics[width=\textwidth]{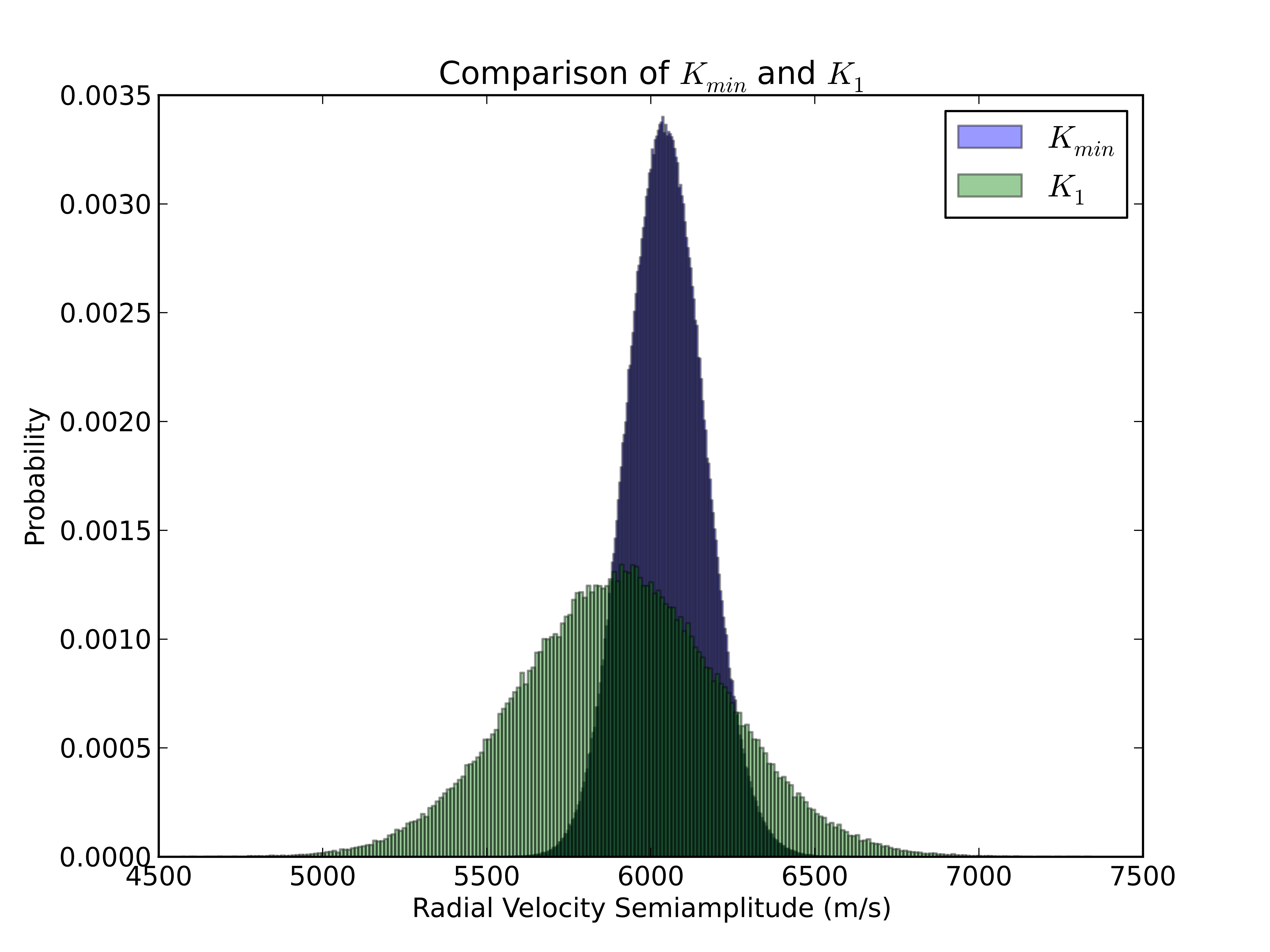}
\caption{A comparison between the PDF of $K_1$ and the bound $K_{min}$. \label{Kcomparison}}
\end{figure*}

We derive a PDF for $K_{min}$, which is compared to that of $K_1$ in Figure \ref{Kcomparison}. These PDFs are consistent with $K_1 = K_{min}$, which is equivalent to $\beta_{I} = 0$. We can consider the contribution of the secondary to the combined luminosity more quantitatively by manipulating Equations \ref{aaleph} and \ref{kepler} a different way to get:

\begin{equation}
\beta_{I}\left(\frac{m_1 + m_2}{m_2}\right) = 1 - \frac{2\pi\sin i}{P\sqrt{1 - e^{2}}}\frac{\aleph}{\varpi}\textrm{AU}\frac{1}{K_1}
\end{equation}

\begin{figure*}[ht]
\centering
\includegraphics[width=\textwidth]{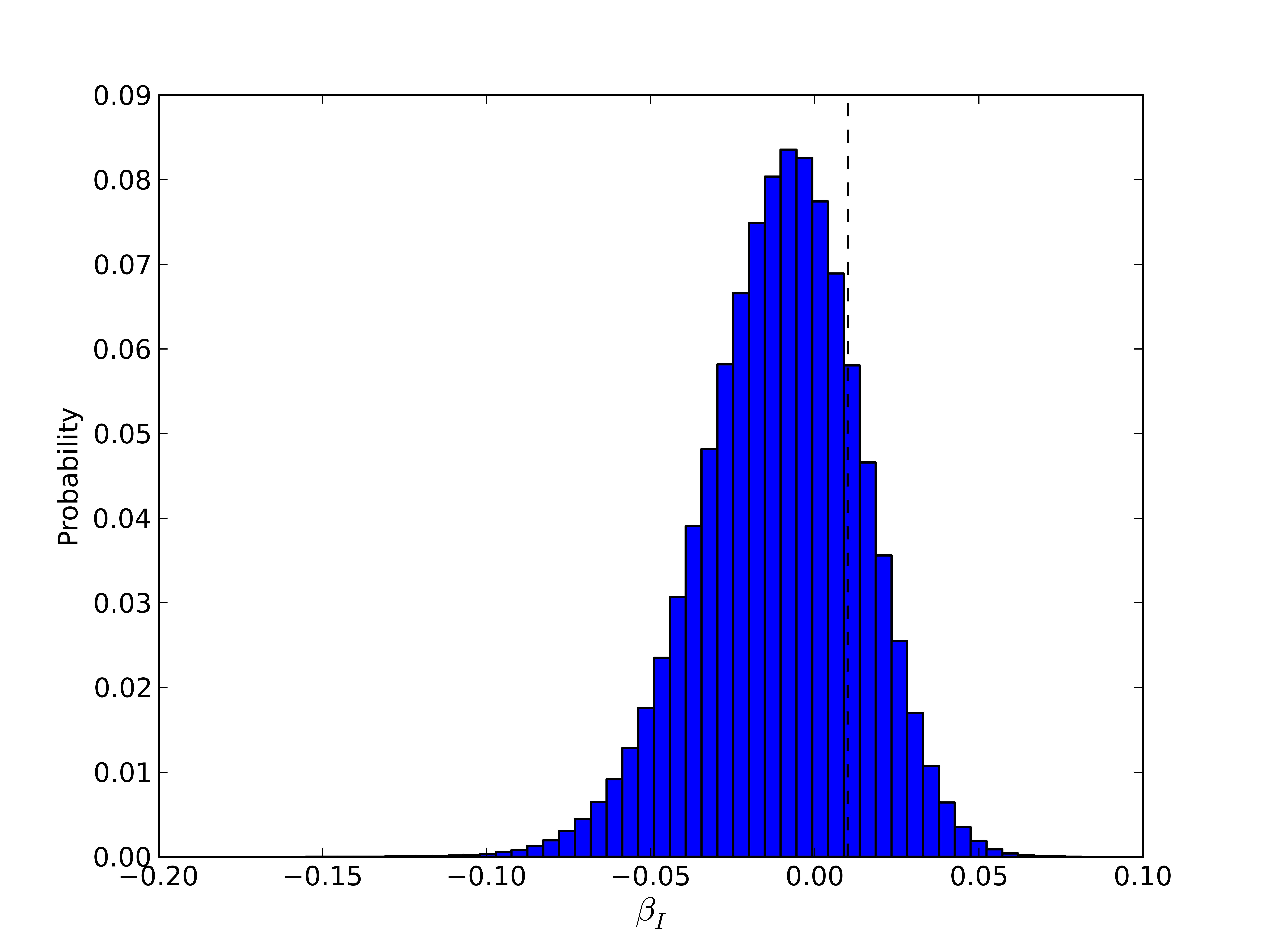}
\caption{The derived probability density function of $\beta_{I}$ with $m_1 = 0.10 M_\odot$. The vertical line denotes $\beta_{I} = 0.01$. \label{betapdf}}
\end{figure*}

For a given value of $m_1$ we have a PDF for $m_2$, which we can then use to place bounds on $\beta_{I}$ itself. For example, we can determine the likelihood that the secondary contributes negligibly to the combined light, say $\beta_{I} \leq 0.01$. We found that the likelihood that $\beta_{I}$ falls below 0.01 is above 80\% for primary masses in the $0.08 M_\odot - 0.30 M_\odot$ range. A sample PDF for $\beta_{I}$ with $m_1 = 0.10 M_\odot$ can be seen in Figure \ref{betapdf}. Based on these results, we assume $\beta_I = 0$ in the subsequent analysis.

We note that these assumptions do not allow us to fully solve the system for the individual component masses. While we had a system of three equations (\ref{aaleph}, \ref{kepler}, \ref{mass}) in four unknowns ($a_1, \beta_{I}, m_1, m_2$), setting $\beta_I = 0$ makes Equation \ref{aaleph} no longer depend on the masses, leaving us with two equations in three unknowns. We do have $a_1 = \aleph/\varpi$, which allows us to derive a precise estimate for $a_1$ from the astrometry alone, giving us $a_1 = 3.22 \pm 0.09 \times 10^{-1}$ AU to 90\% confidence. This is in good agreement with values derived from Equation \ref{kepler}, though since that equation for $a_1$ depends on the $K_1$ calculated from only four RV measurements, it is a weak constraint.

From here, we turn to the photometric data on the combined light of LSR1610AB for additional information on the plausibility of different primary masses. Though  empirical mass-luminosity relations for old, low-mass stars are not well-tested, there are a few lines of evidence we can follow to try to build a consistent picture of the stellar characteristics of LSR1610AB. We preferentially use mass-luminosity relations over mass-color relations since color changes in such low mass stars are confounded by the systematic emergence of broad molecular absorption features at low effective temperatures. 

For our photometric analysis we include apparent magnitude data from the Sloan Digital Sky Survey \citep{aihara}, the Two Micron All Sky Survey \citep{skrutskie}, and observations reported in \citet{dahn}. We must apply the distance modulus $\mu$ to translate these apparent magnitudes into absolute magnitudes:

\begin{equation}
\mathcal{M} - M = \mu = - 5(1 + \log_{10} p) \label{modulus}
\end{equation}

\noindent where $M$ is the absolute magnitude, $\mathcal{M}$ is the apparent magnitude, and $p$ is the parallax in arcseconds. From our 90\% confidence interval on parallax, we can calculate $\mu  = 2.56 \pm 0.02$.

As a first test of plausible primary masses, we use the empirical mass-luminosity relation for very low mass stars set out by \citet{delfosse} in the form of fourth-order polynomial functions for $m(M_\textrm{V})$, $m(M_\textrm{J})$, $m(M_\textrm{H})$, and $m(M_\textrm{K})$. These relations must be used cautiously for our purposes, since benchmark stars that form the basis of these relations are all more massive than about $0.1 M_\odot$, and because we do not know the value of $\beta$ in any of the observing bands. However, these relations are still very useful for estimating masses, and predicted masses are shown in Table \ref{ml}. 

\begin{figure*}[ht]
\centering
\includegraphics[width=\textwidth]{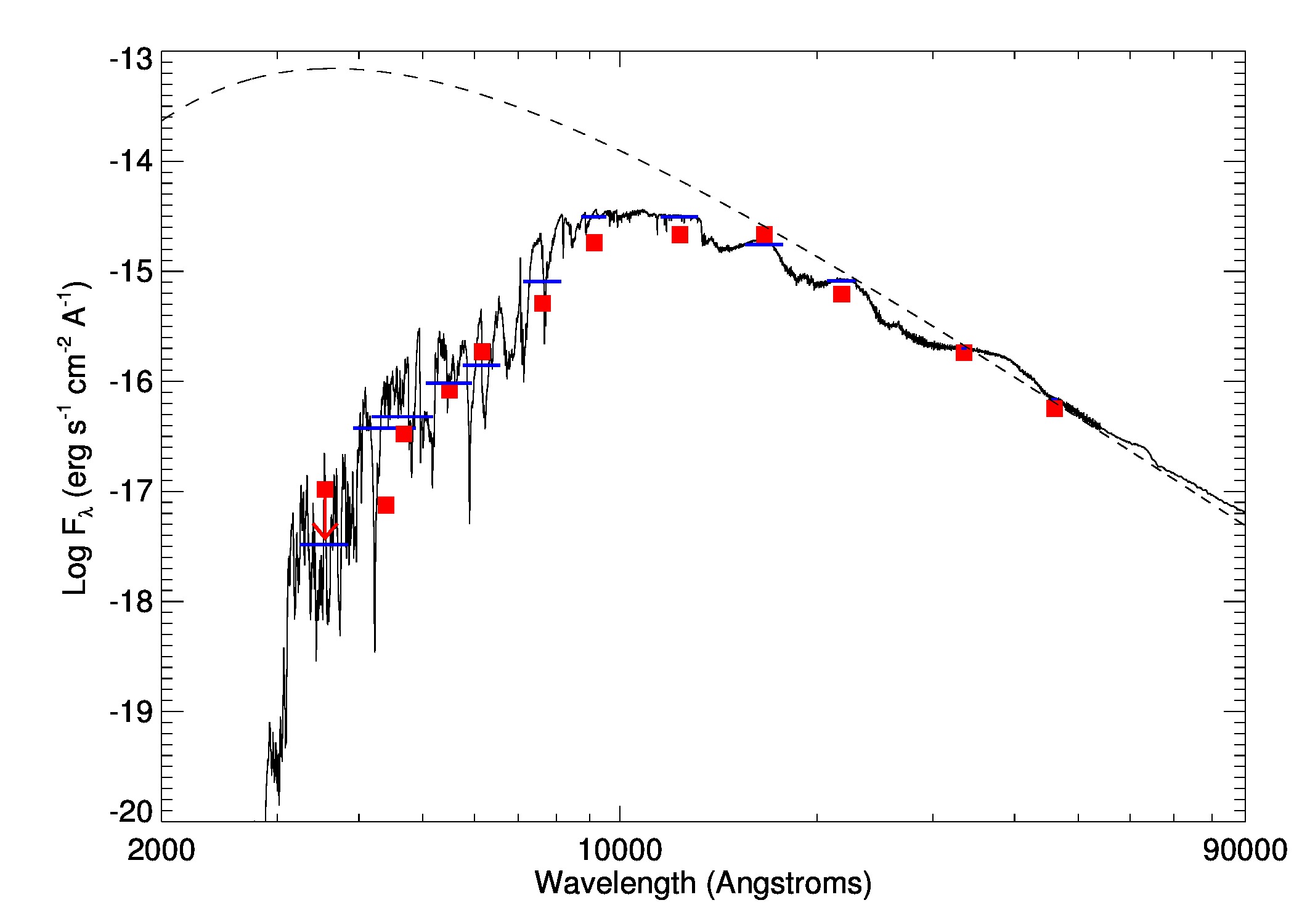}
\caption{Predicted Spectral Energy Distribution from a BT-Settl model \citep{allard}. The model is for metallicity of [Fe/H]=-1, a mass of $\sim 0.1 M_\odot$, a temperature of 2800K, and log(g)=5.0 at a distance of 32.5 pc. Red squares are LSR1610 apparent magnitudes from \citet{dahn}, \citet{aihara}, Skrutskie et al. (2006), and \citet{kirkpatrick}. The shortest wavelength point is a 5$\sigma$ upper limit from SDSS. In all other cases, the photometric errors are smaller than the points. Blue horizontal bars indicate the average flux of the BT-Settl model in each band. The dashed line represents the blackbody SED of a hypothetical old, extremely low mass Helium core white dwarf with log(g)=6.1, T$_{\rm{eff}}=7900$~K, and M=0.155~M$_{\sun}$ based on the evolutionary models presented in \citet{althaus}. An old, low-mass white dwarf companion would easily be detected in the blue photometric bands.\label{sed}}
\end{figure*}

\begin{figure*}[ht]
\centering
\includegraphics[width=\textwidth]{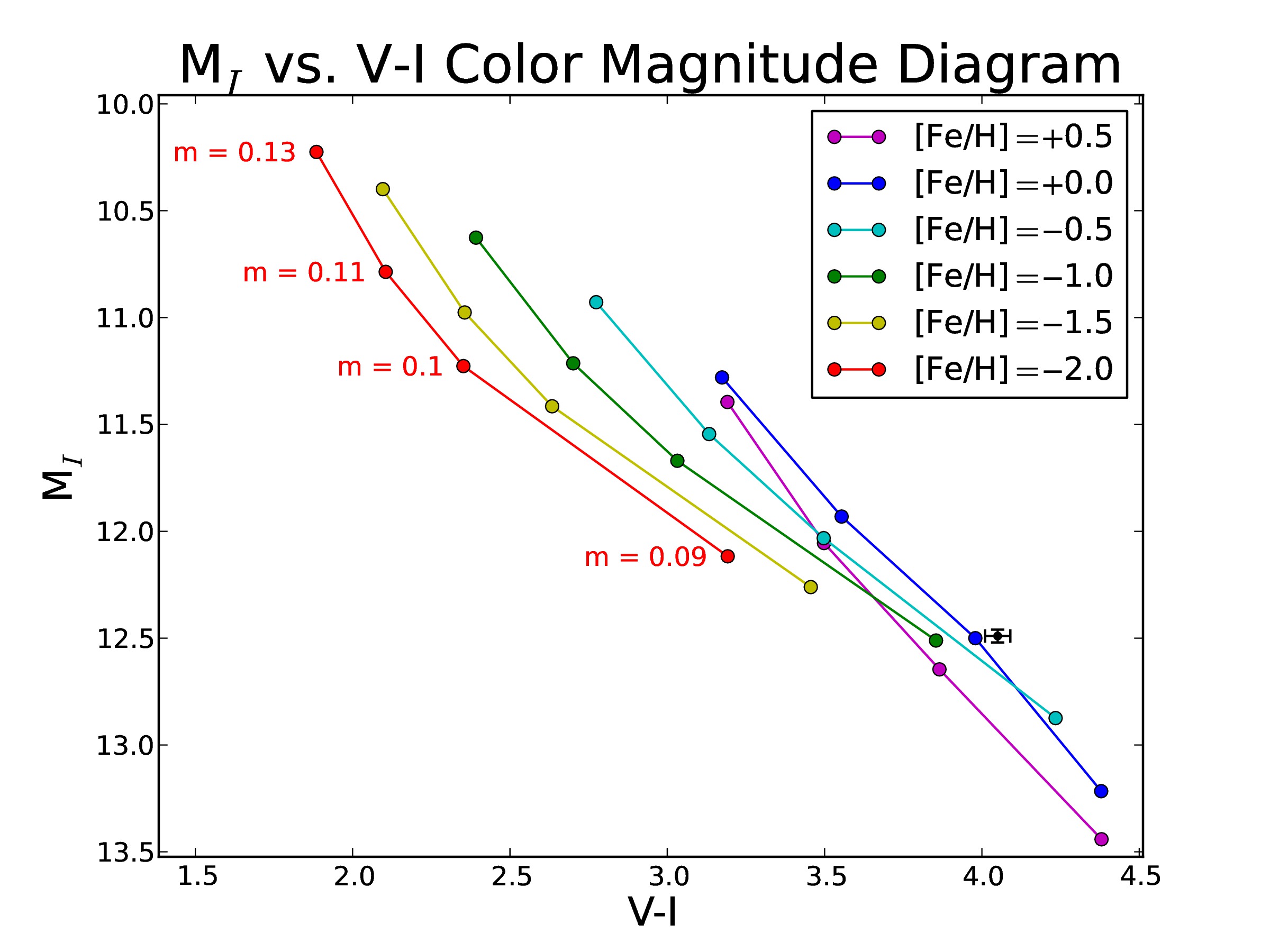}
\caption{Predicted V-I Color vs. I-band absolute magnitude for four different masses of stars in six different metallicities from 5 Gyr BT-Settl models \citep{allard}. The same mass points are plotted for each metallicity, though only one set is labeled. The black square is the position of LSR1610A. A bias in the relative to absolute parallax correction would move the data point downward, though such a bias is expected to be much less than 0.1 magnitude.  \label{vicolor}}
\end{figure*}

We also consider isochrones for low-mass stars and brown dwarfs derived from BT-Settl stellar models \citep{allard}. Since we have taken $\beta_I = 0$, our combined absolute magnitude of LSR1610AB in the I-band is equivalent to the absolute magnitude of LSR1610A in the I-band. So we have $M_{\textrm{I},A} = 12.49 \pm 0.03$ \citep{dahn}. In order to predict the primary mass using the BT-Settl model we need this $M_{\textrm{I},A}$ value and some notion of the metallicity of the primary (a sample SED from a BT-Settl model for $m_1 \sim 0.1 M_\odot$ can be seen in Figure \ref{sed}). We use the $M_\textrm{I}$ vs. V-I color-magnitude relation to infer the plausible metallicities of LSR1610A (see Figure \ref{vicolor}) from 5 Gyr isochrones, as the age is not well known and the available 5 Gyr isochrones cover a large mass range. We find that metallicities in the -1.0 to 0.0 range are broadly consistent with the color and luminosity of LSR1610A, and consider predicted primary masses from BT-Settl isochrones for these metallicities. Interpolated values appear in Table \ref{metalmass}. 

\noindent In summary, three lines of evidence point to a value of $m_1 \sim 0.09-0.10 \ M_\odot$: the astrometric and radial velocity fits, the mass-luminosity relations, and the available isochrones.

\section{\large Discussion and Conclusions}

Based on a combined fit to astrometric and radial velocity observations of LSR1610AB, and comparisons of photometric data to theoretical stellar models, our analysis suggests a most probable primary mass of 0.09-0.10 $M_\odot$. Combining this with the results of our MCMC analysis, we estimate a secondary mass of 0.06 to 0.075 $M_\odot$ with the secondary being essentially dark in the I-band. Given our constraints on the mass of the secondary, a compact or exotic object, such as an old, very-low-mass white dwarf (e.g. \citealt{kilic}) is ruled out. In Figure \ref{sed} we show the broad-band SED of LSR1610AB along with a BT-Settl model for an [Fe/H]=-1.0, 0.1${M_{\Sun}}$ star and an old, very low mass white dwarf model from \cite{althaus}. Given the distance of LSR1610AB, such a white dwarf companion would easily be detected in B- and $u$-band photometry.

Our orbital solution also rules out the possibility of LSR1610 being an eclipsing binary.  As both the primary and secondary are low-mass and therefore have small radii ($\sim R_{jup}$), the comparatively high value of $a_1$ ($\sim 700 \ R_{jup}$) greatly decreases the range of inclination angles from which a transit would be seen, and contributes to an impact parameter of $b > 100$. Transits would only be seen by observers with $i \sim 90\degree \pm{0.04}\degree$, whereas our $1\sigma$ upper bound on $i$ is $85.2\degree$ \citep{winn}.

While the quality of the fit to the astrometric and radial velocity data is quite good, the small number of radial velocity data points ($N_{RV}$ = 4) has widened our bounds on $K_1$, weakening our constraints on the masses of the component stars. However, it would be possible to completely characterize the system from a single photometric measurement of the system \textit{resolved} at apoapsis. Based on our MCMC, an apoapsis should occur on JD 2458097 $\pm$ 12 (2015 May 05 $\pm$ 12 days) and every 633.0 $\pm$ 1.7 days thereafter. Resolving the components and measuring their angular separation would allow us to solve for the component masses through the relation:

\begin{equation}
d_{ap} = (1 + e) \aleph \frac{m_1 + m_2}{m_2} = (1 + e) a
\end{equation}

\noindent where $a = a_1\varpi + a_2\varpi$, and the angular factors have been omitted since $\sin i \sim \sin\Omega \sim 1$. Measuring $d_{ap}$ gives us $a$, and since we already have $a_1$, we could calculate the mass ratio $m_2/m_1$. Together with Equation \ref{x2}, this would give us the component masses.

We can also determine the likely apoapsis angular separation from the analysis in Section 4, which gives $d_{ap}$ is $\sim$ 34 milliarcseconds. This is close to current technological limits. A high-spatial-resolution, high sensitivity observation at the right time should help us solve this system once and for all.

\section*{Acknowledgements}

The authors would like to thank an anonymous referee for a careful reading of the manuscript and his or her many helpful comments and suggestions, which served to signifiantly improve the presentation of these results. SCK gratefully acknowledges the support of the Vagelos Challenge Award at the University of Pennsylvania, and CHB acknowledges support through the NASA Nancy Grace Roman Technology Fellowship program. This work was the senior honors thesis of SCK, and SCK would like to wholeheartedly thank the entire University of Pennsylvania physics department for fostering an exceptional learning and research environment. He especially acknowledges Professors Mirjam Cveti\v{c}, Paul Heiney, Justin Khoury, Tom Lubensky, and Hugh H. ``Brig" Williams.

\FloatBarrier

\FloatBarrier

\begin{table}[p]
\caption{LSR1610 Observations \label{data}}
\begin{center}
\begin{tabular}{|l|r|}
\hline
$N_{RV}$ & 4 \\
Timespan & 4.2 yr \\
$N_{RA}$ & 347 \\
$N_{DEC}$ & 347 \\
Timespan & 10.2 yr \\
\hline
\end{tabular}
\end{center}
\end{table}

\begin{table}[p]
\caption{Point estimates and 90\% confidence intervals on the orbital and system parameters derived from the MCMC fits. The parallax value includes the correction described in Section 4. \label{parameters}}
\begin{center}
\begin{tabular}{|l|r|r|}
\hline
Parameter & Point Estimate & 90\% CI \\ \hline
\multirow{2}{*}{Period ($\textrm{days}$)} & \multirow{2}{*}{$633.0$} & \multirow{2}{*}{$\pm 1.7$} \\
 & & \\
\multirow{2}{*}{Eccentricity} & \multirow{2}{*}{$0.417$} & \multirow{2}{*}{$\pm 0.032$} \\
 & & \\
\multirow{2}{*}{Argument of Periastron ($\degree$)} & \multirow{2}{*}{$146.6$} & $+ 6.0$ \\ & & $- 5.7$ \\
\multirow{2}{*}{Time of Periastron ($\textrm{JD} - 2453500$)} & \multirow{2}{*}{$166$} & $+ 9$ \\ & & $- 10$ \\
\multirow{2}{*}{Radial Velocity Semi-amplitude ($\textrm{m\,s}^{-1}$)} & \multirow{2}{*}{$5910$} & \multirow{2}{*}{$\pm 510$}\\ & & \\
\multirow{2}{*}{Photocentric Semi-Major Axis ($\textrm{mas}$)} & \multirow{2}{*}{$9.89$} & \multirow{2}{*}{$\pm 0.25$} \\
 & & \\
\multirow{2}{*}{Inclination Angle ($\degree$)} & \multirow{2}{*}{$83.5$} & \multirow{2}{*}{$\pm 1.7$} \\
 & & \\
\multirow{2}{*}{Longitude of Ascending Node ($\degree$)} & \multirow{2}{*}{$99.5$} & \multirow{2}{*}{$\pm 1.8$} \\
 & & \\
\multirow{2}{*}{Parallax ($\textrm{mas}$)} & \multirow{2}{*}{$30.73$} & \multirow{2}{*}{$\pm 0.34$} \\
 & & \\
\multirow{2}{*}{RA Proper Motion ($\textrm{mas\,yr}^{-1}$)} & \multirow{2}{*}{$-795.42$} & \multirow{2}{*}{$\pm 0.07$} \\
 & & \\
\multirow{2}{*}{Dec Proper Motion ($\textrm{mas\,yr}^{-1}$)} & \multirow{2}{*}{$-1208.40$} & \multirow{2}{*}{$\pm 0.9$} \\
 & & \\
\multirow{2}{*}{Systemic Radial Velocity ($\textrm{km\,s}^{-1}$)} & \multirow{2}{*}{$-98.3$} & \multirow{2}{*}{$\pm 0.3$} \\
 & & \\
\hline
\end{tabular}
\end{center}
\end{table}

\begin{table}[p]
\caption{Goodness of Fit Statistics\label{goodness}}
\begin{center}
\begin{tabular}{|l|r|}
\hline
$\chi_{RV}^{2}$ & 1.3 \\
$\chi_{RA}^{2}$ & 361.5 \\
$\chi_{DEC}^{2}$ & 329.6  \\
$RMS_{RV}$ & 0.26 ($\textrm{km\,s}^{-1}$) \\
$RMS_{RA}$ & 2.3 (mas) \\
$RMS_{DEC}$ & 3.2 (mas)  \\
\hline
\end{tabular}
\end{center}
\end{table}

\begin{table}[p]
\caption{Primary Masses and Associated Secondary Mass Ranges\label{masses}}
\begin{center}
\begin{tabular}{|r|r|r|}
\hline
$m_1$ ($M_\odot$) & $m_2$ ($M_\odot$) Point Estimate & 90\% CI \\ \hline
\multirow{2}{*}{0.08} & \multirow{2}{*}{0.061} & $+0.008$ \\ & & $-0.007$ \\
\multirow{2}{*}{0.09} & \multirow{2}{*}{0.065} & \multirow{2}{*}{$\pm 0.008$} \\ & & \\
\multirow{2}{*}{0.10} & \multirow{2}{*}{0.069} & \multirow{2}{*}{$\pm 0.008$} \\ & & \\
\multirow{2}{*}{0.11} & \multirow{2}{*}{0.072} & $+0.009$ \\ & & $-0.008$\\
\multirow{2}{*}{0.12} & \multirow{2}{*}{0.076} & \multirow{2}{*}{$\pm 0.009$} \\ & & \\
\multirow{2}{*}{0.15} & \multirow{2}{*}{0.086} & \multirow{2}{*}{$\pm 0.010$} \\ & & \\
\multirow{2}{*}{0.20} & \multirow{2}{*}{0.101} & $+0.012$ \\ & & $-0.011$\\
\multirow{2}{*}{0.25} & \multirow{2}{*}{0.115} & $+0.013$ \\ & & $-0.012$\\
\multirow{2}{*}{0.30} & \multirow{2}{*}{0.128} & \multirow{2}{*}{$\pm 0.014$} \\ & & \\ \hline
\end{tabular}
\end{center}
\end{table}

\begin{table}[p]
\caption{Mass Estimates From Delfosse et al.(2000) Mass-Luminosity Relations for Very Low Mass Stars - Assuming $\beta=0$\label{ml}}
\begin{center}
\begin{tabular}{|l|r|r|}
\hline
Band & Absolute & Estimated Primary  \\ 
& Magnitude &  Mass ($M_\odot$) \\ \hline
$M_\textrm{V}$ & $16.54 \pm 0.03$ & 0.098  \\
$M_\textrm{J}$ & $10.35 \pm 0.04$ & 0.099 \\
$M_\textrm{H}$ & $9.76 \pm 0.03$ & 0.091 \\
$M_\textrm{K}$ & $9.46 \pm 0.04$ & 0.094  \\
\hline
\end{tabular}
\end{center}
\end{table}

\begin{table}[p]
\caption{Metallicities and Associated Masses from Mass-Luminosity Relation and Isochrones \label{metalmass}}
\begin{center}
\begin{tabular}{|l|r|r|}
\hline
Primary Metallicity & $m_1$ ($M_\odot$) & Bounds \\ \hline
\multirow{2}{*}{\textbf{$[Fe/H] = 0.0$}} & \multirow{2}{*}{$0.010$} & \multirow{2}{*}{$\pm 0.005$} \\
 & & \\
\multirow{2}{*}{\textbf{$[Fe/H] = -0.5$}} & \multirow{2}{*}{$0.093$} & $+ 0.004$ \\ && $- 0.003$ \\
\multirow{2}{*}{\textbf{$[Fe/H] = -1.0$}} & \multirow{2}{*}{$0.090$} & $+ 0.002$ \\ && $- 0.003$ \\ \hline
\end{tabular}
\end{center}
\end{table}

\clearpage 

\clearpage 

\clearpage 

\clearpage 

\clearpage 

\clearpage 

\clearpage 

\clearpage 

\end{document}